\documentclass{article}

\usepackage{arxiv}

\usepackage[utf8]{inputenc} % allow utf-8 input
\usepackage[T1]{fontenc}    % use 8-bit T1 fonts
\usepackage{hyperref}       % hyperlinks
\usepackage{url}            % simple URL typesetting
\usepackage{booktabs}       % professional-quality tables
\usepackage{amsfonts}       % blackboard math symbols
\usepackage{nicefrac}       % compact symbols for 1/2, etc.
\usepackage{microtype}      % microtypography
\usepackage{lipsum}

\usepackage{bm}

\title{Notation for Subject Answer Analysis}

\author{
  Lucjan Janowski\\
  Department of Telecommunication\\
  AGH University of Science and Technology\\
  Kraków, Poland\\
  \texttt{janowski@kt.agh.edu.pl} \\
 	\And
Jakub Nawała \\
  Department of Telecommunication\\
AGH University of Science and Technology\\
Kraków, Poland\\
\texttt{jnawala@kt.agh.edu.pl} \\
   \And
 Werner Robitza \\
 Audiovisual Technology Group\\
Technische Universität Ilmenau\\
Ilmenau, Germany\\
\texttt{werner.robitza@tu-ilmenau.de} \\
 	\And
 Zhi Li\\
 Netflix\\
 \texttt{zli@netflix.com} \\
 	\And
 Luk\'{a}\v{s} Krasula \\
 Arm Ltd. \\
 Loughborough, United Kingdom \\
 \texttt{lukas.krasula@arm.com} \\
 	\And
 Krzysztof Rusek \\
  Department of Telecommunication\\
AGH University of Science and Technology\\
Kraków, Poland\\
\texttt{krusek@kt.agh.edu.pl} \\
}

\begin{document}
\maketitle

\begin{abstract}
It is believed that consistent notation helps the research community in many ways. First and foremost, it provides a consistent interface of communication. Subjective experiments described according to uniform rules are easier to understand and analyze. Additionally, a comparison of various results is less complicated. In this publication we describe notation proposed by VQEG (Video Quality Expert Group) working group SAM (Statistical Analysis and Methods).
\end{abstract}

% keywords can be removed

\section{The Proposed Notation}

We start from the notation itself, just to make it easy to find and use it. At the end of this paper more detail reason for this notation is given.

The general concept is based on two assumptions:
\begin{itemize}
	\item capital letters as random variables,
	\item Greek letters to describe model parameters
\end{itemize}

The interpretation of letters is:
\begin{itemize}
	\item $u$ as a single subject answer,
	\item $\psi$ (psi) as a true quality,
	\item $\Delta$ (Delta) as a subject bias,
	\item $\upsilon$ (upsilon) as a standard deviation related with a given subject,
	\item $\phi$ (phi) as a standard deviation related with a given PVS
(Processed Video Sequence, a specific sequence shown to a subject), and
	\item $\rho$ (rho) as a standard deviation related with a given SRC (SRC
is a source sequence from which different PVSs are generated).
\end{itemize}

What is more, we suggest a following set of indices:
\begin{itemize}
	\item $i$ for a subject
	\item $j$ for a PVS,
	\item $k$ for an SRC,
	\item $r$ for a repetition,
	\item $o$ for an order, and
	\item $h$ for an HRC (Hypothetical Reference Circuit, a way a particular SRC is treated to generate PVS).
\end{itemize}

For indices which are related to each other, like SRC ($k$) and PVS ($j$), we use the notation $k:\boldsymbol{k}(j)=k$ It uses the $\boldsymbol{k}()$ function (accepting a PVS and returning an SRC) to map the PVS to the related SRC. A similar function can be found for an HRC, $h: \boldsymbol{h}(j) = h$.

Importantly, we advice to skip all indices not appropriate for a given context (e.g. do not use $r$ if there are no repetitions in your subjective test).

We also propose to place a horizontal bar above a parameter if an averaging operation is done. For example, an average value (over all subjects) of a sequence
quality for some PVS $j$ is denoted as $\bar{u}_j$.

Similarly, if an estimation of some parameter is done, it should be marked by a hat placed above it. For example, an estimation of a standard deviation related with some subject $i$
is given as $\hat{\upsilon}_i$.

A single raw opinion score from a subjective test without any repetitions is denoted as $u_ij$ (small $u$). MOS for PVS $j$ is now given as $\bar{u}_j$. An estimation of a true quality based on the subject model selected (called the adjusted MOS) is denoted as $\bar{psi}_j$.

Importantly, the adjusted MOS is just one of many possible estimations of the true quality. The most obvious other one is a classic MOS. Thus, both the adjusted MOS and MOS can be called the estimated true quality, However, it is recommended to use specific terms instead, either the MOS or adjusted MOS.

If we use a specific answer we should use letter $s$ (like score), where $s \in (1, \cdots, S)$. Probability of any specific answer is $P(U_{ij} = s)$.

\subsection{Examples}

The model proposed by Li and Bampis \cite{Li2017recover} takes the following form:
\begin{equation}
U_{ij}=\psi_j+\Delta_i+\upsilon_i X + \rho_{k:\boldsymbol{k}(j)=k}Y\textrm{,}
\end{equation}
where: $U_{ij}$ is a random variable describing raw opinion scores, $\boldsymbol{k}(j)$ is a mapping function that returns the SRC $k$ from which the PVS $j$ was created, and $X, Y \sim \mathcal{N}(0,1)$.

The model proposed by Janowski and Pinson \cite{janowski2015accuracy} is now as follows:
\begin{equation}
U_{ij}=\psi_j + \Delta_i + \upsilon_i X + \phi_j Y
\end{equation}

If an experiment analysis takes into account an order of answers, a special notation is used. For example, to express a calculation of the subject bias for the first and last 25 sequences (in an experiment with 200 PVSs shown to one subject), a subject bias estimation is described by two following equations:
\begin{equation}
\bar{\Delta}_{i,\mathrm{start}}=\frac{1}{25}\sum_{o=1}^{25}(u_{ijo}-\hat{\psi}_j)
\end{equation}
\begin{equation}
\bar{\Delta}_{i,\mathrm{end}}=\frac{1}{25}\sum_{o=176}^{200}(u_{ijo}-\hat{\psi}_j)
\end{equation}

\section{Justification}

Each and every letter used in the notation proposed has some justification.
Following section lists those justifications along with corresponding letters.

\begin{itemize}
	\item $u$ as a single subject answer - taken from the BT.500 Recommendation \cite{ITURBT500}
	\item $\psi$ (psi) as a true quality - corresponds to some unknown data, which is usually denoted like that.
	\item $\Delta$ (Delta) as a subject bias - represents a shift, which is usually denoted like that.
	\item $\upsilon$ (upsilon) as a standard deviation related with a given subject - a Greek letter ``u.'' It corresponds to the ``u'' letter appearing in words User and sUbject.
	\item $\phi$ (phi) as a standard deviation related with a given PVS - resembles a Greek letter $\pi$ (usage of would look confusing). It corresponds to the first letter in a word PVS.
	\item $\rho$ (rho) as a standard deviation related with a given SRC - a Greek letter ``r.'' It corresponds to the ``r'' letter in a word SRC.
	\item Indices $i$ (subject), $j$ (PVS), $k$ (SRC), $r$ (repetition) are taken from the BT.500 \cite{ITURBT500}. 
	\item Index $o$ (order) and index $h$ (HRC) are used as they correspond to letters ``o'' and ``h'' appearing in words order and HRC respectively.
\end{itemize}

\section{Notation Literature Review}

In this section we cite the existing notation dividing it to recommendations or standards and publications.

\subsection{Standards}

With a standard the publication date can be misleading since the notation can be proposed in a previous version.

\subsubsection{P.1401, 07.2012}

This recommendation is focused on the subjective data analysis so it should be the best source of information. The notation for a single answer is:
$$
v_{j,l,k}
$$
where $v$ is a subject opinion, $j$ is condition, $k$ is listener, $l$ is talker. For video case it would be $j$ is HRC, $k$ is subject, and $l$ is SRC. In different part of document PVS is marked by $i$.

\subsubsection{BT.500-13, 01.2012}

This is probably the most cited recommendation in the video quality community. On page 36 we can find:
$$
u_{ijkr}
$$
where $i$ is observer, $j$ is test condition, $k$ is sequence/image, and $r$ is repetition. For our notation it would be $i$ - subject, $j$ - HRC, $k$ - SRC, and $r$ - repetition.

On page 39, the notation is changed in the case of multiple answers per PVS for:
$$
u_{njklr}
$$
where $n$ is observer, $j$ is a number of voting time window within single PVS voting period,  $k$ is test condition, $l$ is sequence, and $r$ is repetition. Sadly it seems that BT.500-13 is not consistent with itself.

\subsection{Publications}

\subsubsection{SOS: The MOS is not enough!, 09.2011}

This is probably one of the first and very important papers about subjective experiments as a measuring tool, but it does not refer to a single answer so there is no notation we could use \cite{HSEorSOS}.
 
\subsubsection{Subject Bias: Introducing a Theoretical User Model, 2014}

The main goal of this publication is to propose subjects’ bias. A single answer is divided to three different factors. The main equation is:
$$
o_{ij}=\psi_j + \Delta_i + \epsilon_{ij}
$$
where $o_{ij}$ is single answer from subject $i$ given to PVS $j$, $\psi_j$ is the true quality,  $\Delta_i$ is a subject bias \cite{SubjectBias}. 

\subsubsection{The Accuracy of Subjects in a Quality Experiment: A Theoretical Subject Model, 2015}

The main goal of this paper is to extend the previous model. With the extension the error term was divided into two parts. Error introduced by subject and PVS. The most general equation is (16):
$$
o_{ijr}=\psi_j+\Delta_i+\alpha_iX+\beta_jY
$$

where $o_{ijr}$ is a single answer from subject $i$ given to PVS $j$ with repetition $r$, note that for single answer $r=1$, to this equation describes single answer as well as multiple answers per PVS, $\psi_j$ is true quality, $\Delta_i$ is a subject bias, $\alpha_i$ is standard deviation introduced by subject $i$, $\beta_j$ is standard deviation introduced by PVS $j$, and both $X$ and $Y$ are independent with the standard normal distribution $X,Y\sim \mathcal{N}(0,1)$ \cite{janowski2015accuracy}.
 
\subsubsection{Recover Subjective Quality Scores from Noisy Measurements, 2017}

The main goal of this paper is to propose subject model with a different error term than proposed in \cite{janowski2015accuracy}. Also the estimation method is improved. The main equation is (1):
$$
X_{e,s}=x_e+B_{e,s}+A_{e,s}
$$
where $X_{e,s}$ is subject opinion for PVS $e$ given by subject $s$, $x_e$ represents the quality of impaired video (PVS) $e$ perceived by an average viewer, $B_{e,s}\sim \mathcal{N}(b_s, v_s)$ are i.i.d. Gaussian variables representing the factor of subject $s$, and $A_{e,s}\sim\mathcal{N}(0,a_{c:\boldsymbol{c}(e)=c})$,  are i.i.d. Gaussian variables representing the factor of video content (SRC) $c$ (i.e., the content that $e$ corresponds to). The parameters $b_s$ and $v_s$ represent the bias (i.e., mean) and inconsistency (i.e., standard deviation) of subject . The parameter  represents the ambiguity (i.e., standard deviation) of content $c$.

\subsubsection{Confidence Interval Estimators for MOS Values, 2018}

This paper analyzes with details the confidence interval obtained by different methods. Just at the beginning it introduces notation:
$Y_x$ - random variable of user ratings for test condition (TC) $x$
$k$ - users rate on a discrete $k$-point rating scale from $1, \cdots, k$
$n$ - number of users rating the test condition
$m$ - number of TC
$r$ - number of simulation runs
$y_{u,x,i}$ - sampled user rating for user $u$, TC $x$ and simulation run $i$
$\hat{Y}_{x,i}$ - MOS, i.e. sample mean over user ratings, for TC $x$ and run $i$
$\gamma$ - confidence level
$\alpha$ - significance level, e.g. for $\alpha=0.05$ we have $\gamma = 1 - \alpha$ \cite{Hossfeld2018}.

The paper presents a different method for generating subjects answers. It is generated by binomial distribution. So there is no simple relation from the previous notation to this paper notation. 

\bibliographystyle{unsrt}  
\bibliography{references} 

\end{document}